\documentclass[epj]{svjour}
\usepackage{latexsym}
\usepackage{graphicx}
\usepackage{latexsym}
\usepackage{color}

\begin{document}
\title{Polymer and surface roughness effects on the drag crisis for falling spheres}
\author{Nicolas Lyotard\inst{1} \and Woodrow L. Shew\inst{1} \and Lyd\'{e}ric Bocquet\inst{2} \and Jean-Fran\c{c}ois Pinton\inst{1}}
\institute{Laboratoire de Physique de l'\'{E}cole Normale
Sup\'{e}rieure de Lyon, CNRS UMR5672, \\
46 all\'ee d'Italie F-69007 Lyon, France \and Laboratoire de
Physique de la Mati\`{e}re Condens\'{e}e et Nanostructures\\
  Universit\'{e} Lyon I, 43 Bd du 11 Novembre 1918,
  69622 Villeurbanne}
\date{Received: date / Revised version: date}
%
\abstract{ We make time resolved velocity measurements of steel
spheres in free fall through liquid using a continuous ultrasound
technique. We explore two different ways to induce large changes
in drag on the spheres: 1) a small quantity of viscoelastic
polymer added to water and 2) altering the surface of the sphere.
Low concentration polymer solutions and/or a pattern of grooves in
the sphere surface induce an early drag crisis, which may reduce
drag by more than 50 percent compared to smooth spheres in pure
water. On the other hand, random surface roughness and/or high
concentration polymer solutions reduce drag progressively and
suppress the drag crisis.  We also present a qualititative argument
which ties the drag reduction observed in low concentration
polymer solutions to the Weissenberg number and normal stress difference.
}
\maketitle

\section{Introduction}
\label{intro} Reduction of drag in turbulent flows due to a small
quantity of viscoelastic polymer added to the fluid has been the
subject of intense research for more than 50 years
(e.g.~\cite{lumley,berman}). For example, the addition of as
little as 5 parts per million (ppm) of polyacrylamide to turbulent
pipe flow can result in an increase in flow speed of 80 percent
for a given imposed pressure drop~\cite{toms}.  Similar flows with
rough or structured wall surfaces have also been shown to exhibit
reduction in drag (e.g.~\cite{hanratty,deutsch}). Our experiments
address drag reduction by similar mechanisms for bluff bodies,
which has received far less attention in spite of the potential
impact on a broad range of phenomena and applications (aircraft,
underwater vehicles and ballistics, predicting hail damage, sports
ball aerodynamics, fuel pellets, etc.).

The aim of our work is to explore the influence of polymer
additives in the fluid as well as sphere surface structure on the
drag experienced by free falling spheres. Before we review the
literature on these topics, let us first recall the main
characteristics and terminology of high Reynolds number flow
around spheres. (Reynolds number is defined as $Re=UD/\nu$ where
$U$ is sphere speed, $D$ is sphere diameter, and $\nu$ is the
kinematic viscosity of the fluid.)  In the range $10^4 < Re <
10^7$, two basic phenomena are responsible for the most prominent
flow features: flow separation and the transition to turbulence in
the sphere boundary layer.  For $200 < Re < Re^*_{\rm w}\approx
3\times 10^5$ flow separation occurs.  (The $w$ subscript
distinguishes the value for smooth spheres in pure water from the
different cases discussed later.) In this regime, laminar flow
extends from the upstream stagnation point to slightly downstream
of the flow separation point, i.e. the turbulence develops {\it
downstream} from the separation point. In contrast, at $Re$ just
above $Re^*_{\rm w}$, the boundary layer becomes turbulent {\it
upstream} of the flow separation point. The resulting change in
the velocity profile abruptly moves the separation point
downstream.  Since the drag on the sphere is dominated by pressure
drag (form drag), this jump in the separation location results in
a severe drop in drag, the so-called {\it drag
crisis}~\cite{achenbach1,taneda,maxworthy,squires}. In this range
of $Re$, friction drag contributes not more than 12~percent to the
total drag on a smooth sphere~\cite{achenbach1}. Although
indirectly, our investigation is essentially exploring the effects
of polymer additives and sphere surface structure on the dynamics
of boundary layer separation and transition to turbulence.

We now briefly review studies which directly address these issues.
Both Ruszczycky~\cite{rusz} and D. A. White~\cite{dawhite}
measured drag on a falling sphere in aqueous polymer solutions at
$Re < Re^*_{\rm w}$. Ruszczycky studied relatively high
concentrations between 2500 and 15000~ppm (by weight) of
poly(ethylene oxide) ($4\times 10^6$ molecular weight (MW)) and
guar gum (unknown MW) for a range of sphere sizes from 9.5 to
25.4~mm in diameter. Maximum drag reduction of 28 percent was
found for a 25.4~mm sphere in 5000~ppm guar gum solution.  For
higher concentrations (15000~ppm) the drag was found to increase
compared to water, probably because such high concentrations tend
to be rather viscous.  D. A. White used the same polymer at
smaller concentrations with a similar range of sphere sizes and
found a 45 percent maximum reduction in drag for a 75 ppm
solution.   A. White~\cite{awhite} and more recently Watanabe et
al.~\cite{watanabe} investigated a range of $Re$ spanning the drag
crisis.   Their work suggests that at polymer concentrations above
about 30~ppm, the drag crisis is replaced by a gradual decrease in
drag which manifests as drag reduction for $Re < Re^*_{\rm w}$ and
drag enhancement for supercritical $Re > Re^*_{\rm w}$. This is
consistent with the observations of D. A. White and Ruszczycky
below the drag crisis, as well as water tunnel measurements with
circular cylinders~\cite{sarpkaya}. At smaller polymer
concentrations (5 to 10 ppm) the situation is less clear. A.
White's measurements show erratic variation of drag as $Re$ is
increased, while Watanabe et al. report no change in behavior
compared to water. Cylinder studies, in contrast, show a more
sharply defined drag crisis at low polymer
concentration~\cite{sarpkaya}. One of the goals of our work is to
better understand the nature of low concentration polymer effects
near the drag crisis.

Concerning free falling rough spheres, to our knowledge, only one
experimental work exists in the literature. In this short,
qualitative study, A. White explored the combined effects of
surface roughness and polymer additives \cite{awhite2}. He found
that roughening the sphere surface shifts the wake separation
point downstream, reducing drag, but with both a rough surface and
polymers added the separation point shifts back upstream,
increasing drag.  Our observations add to White's intriguing
results.

Wind tunnel measurements for both spheres \cite{achenbach2} and
cylinders \cite{nakamura} indicate that the drag crisis is shifted
to lower $Re$ when the surface is roughened. The roughened surface
triggers an early transition to turbulence in the boundary layer.
Golf balls are made with surface dimples in order to reduce drag
by a very similar mechanism~\cite{golf}.  Furthermore, Maxworthy
showed that adding a trip on the upstream surface of a smooth
sphere induces a turbulent boundary layer and early drag
crisis~\cite{maxworthy}.  We are aware of no fixed sphere studies
addressing roughness and polymer effects together. We add a note
of caution to the reader that fixed (wind tunnel) and free-falling
spheres may not behave the same. The first case corresponds to a
fixed velocity of the upstream flow, while the second corresponds
to a constant force driving the motion.  Unlike the fixed sphere,
a falling sphere cannot exist in a steady state with $Re$ very
close to $Re^*_{\rm w}$; it is not a stable solution. Furthermore,
even at terminal fall speed the wake is never truly steady. It is
dynamically active with long-lived non-axisymmetric spatial
structure. As a result, the ``terminal" fall velocity of a sphere
fluctuates in both direction and magnitude, which may lead to
small discrepancies in comparing to wind tunnel data or to other
free-fall experiments.

This paper is organized as follows.  The next section presents the experimental procedures and equipment.  In section~\ref{results} we present our measurements for the free fall of smooth or roughened spheres in water and in solutions
containing small polymer amounts. We discuss our results in terms
of changes in drag with varying Reynolds number, polymer
concentration and surface conditions. In the last section, we link our results at low polymer concentrations to the effects of  a coil-stretch transition~\cite{deGennes}.

\begin{figure}[t!]
\center
\centerline{\includegraphics[width=8.5cm]{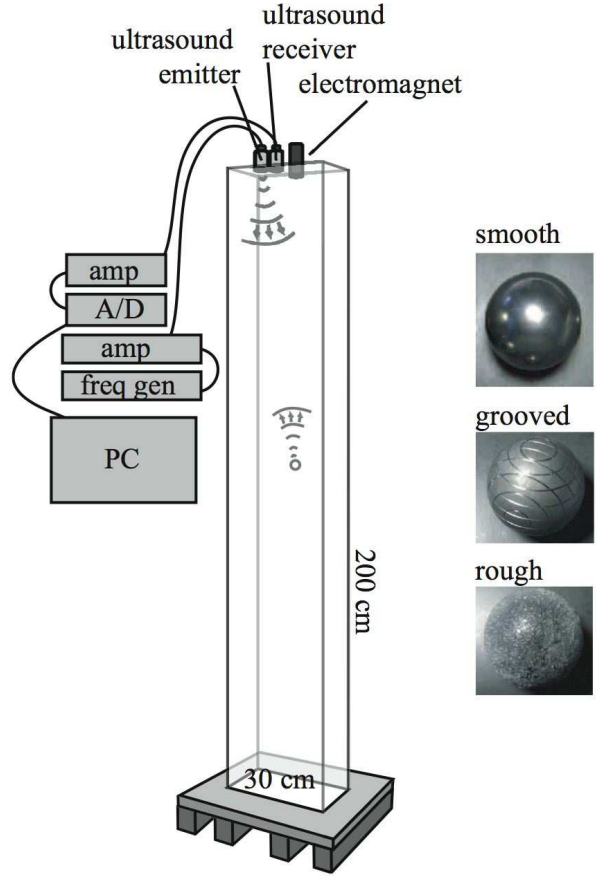}}
\vspace{0.5cm} \caption{SETUP:  the vertical velocity of falling steel spheres is measured with an ultrasound device. The fluid is tap water, pure or with small amounts of pomymer additives. Smooth, grooved, and
rough spheres are tested. 
}\label{setup}
\end{figure}

\section{Measurement system and technique}
We measure the fall velocity of steel spheres (ball bearings with
density $\rho=7.8$~g/cm$^3$) with diameters ranging from 3~mm to
80~mm.  Two types of sphere surfaces are investigated in addition
to the polished smooth surface (see photos in fig.~\ref{setup}).
The first type, grooved spheres, have have a regular pattern of
grooves machined into the surface. The grooves are 500~$\mu$m
deep, 1~mm wide.  The second type corresponds to roughened
surface, produced either by sanding the smooth polished spheres or
by gluing onto the surface a single layer beads. In the first
case, changes in the surface height are of the order of 10
microns. In the second case, we have used spherical glass beads
700~$\mu$m in diameter.  The fluid vessel is 2~m tall and 30
$\times$ 30~cm in lateral dimension with walls made of 2~cm thick
acrylic plate. The tank is filled with either tap water or a
dilute aqueous solution of polyacrylamide (MW 5$\times 10^6$,
granulated form, Sigma-Aldrich). The polymer solutions range in
concentration between 5 and 200 ppm by weight. The polymers are
mixed first with 2~liters of water with a magnetic stirrer for at least
8 hours and then mixed with another 180~liters of water for 5 minutes
in the experimental vessel. Tests with colored dye in the fluid
confirm that this procedure effectively mixes the fluid. These
polymer concentrations are in the dilute regime, significantly
below the estimated overlap concentration of 1200 ppm.  The
Weissenberg number $Wi$, defined as the ratio of polymer
relaxation time $\tau_R \sim 10^{-4}$ to flow time scale (see
section~\ref{lyderic} for details), ranges between 0.8 and 2.3.

The spheres are released at the top of the vessel using an
electromagnet.  The speed of the ball is obtained using a
continuous ultrasound technique.  This technique is described in
more detail in previous publications \cite{mordant1,mordant2},
but we briefly describe it here.  One ultrasound transducer
positioned at the top of the vessel emits sound at 2.8~MHz into
the fluid. As the sphere falls it scatters sound at a Doppler
shifted frequency which is measured with a second ultrasound
transducer located near the emitter.  The recorded signal is
processed to recover the vertical component of the sphere
velocity.  The processing entails mixing the recorded signal with
a 2.8~MHz sinusoid, low pass filtering, decimating to a lower
sample rate, and finally using a parametric time-frequency
analysis algorithm (MVA, see ref.~\cite{mordant2}) to recover the
time varying Doppler shifted frequency.  The resulting absolute
precision for the velocity measurement is about 2~cm/s with a
relative precision in mm/s.  With typical fall speeds of
several m/s, this is better than 1\% precision.  At such high
Reynolds numbers ($10^4-10^6$), the flow in the wake contains
significant non-axisymmetric flow structures \cite{taneda}, which
often cause some lateral motion of the sphere. We present data
only from trajectories that remained at least one sphere diameter
away from the vessel walls throughout the fall. Based on studies
of tunnel blockage effects for fixed spheres, we expect that walls
have less than 5\% influence on measured drag coefficients
\cite{achenbach2}.  Furthermore, any wall influence is similar for
the different polymer solutions and sphere surfaces allowing for
meaningful {\it comparisons} between the different cases.

\section{Experimental results}\label{results}
In this section, we present our observations in the form of either
drag coefficient estimates or velocity time series.  Each
presented measurement is the result of averaging over several
trajectories under the same conditions. We find that each drop is
reproducible up to instantaneous differences of a several percent.
We first discuss our measurements of smooth spheres falling in
water, which provide a baseline for comparisons to the results
from our experiments with polymer solutions and rough spheres.
Next, we present measurements of smooth sphere behavior in polymer
solutions.  Then we explore the consequences of surface grooves or
roughness in water. And finally, we address the combined case of
altered-surface spheres in polymer solutions.

\begin{figure}[t!]
\center
\centerline{\includegraphics[width=7.5cm]{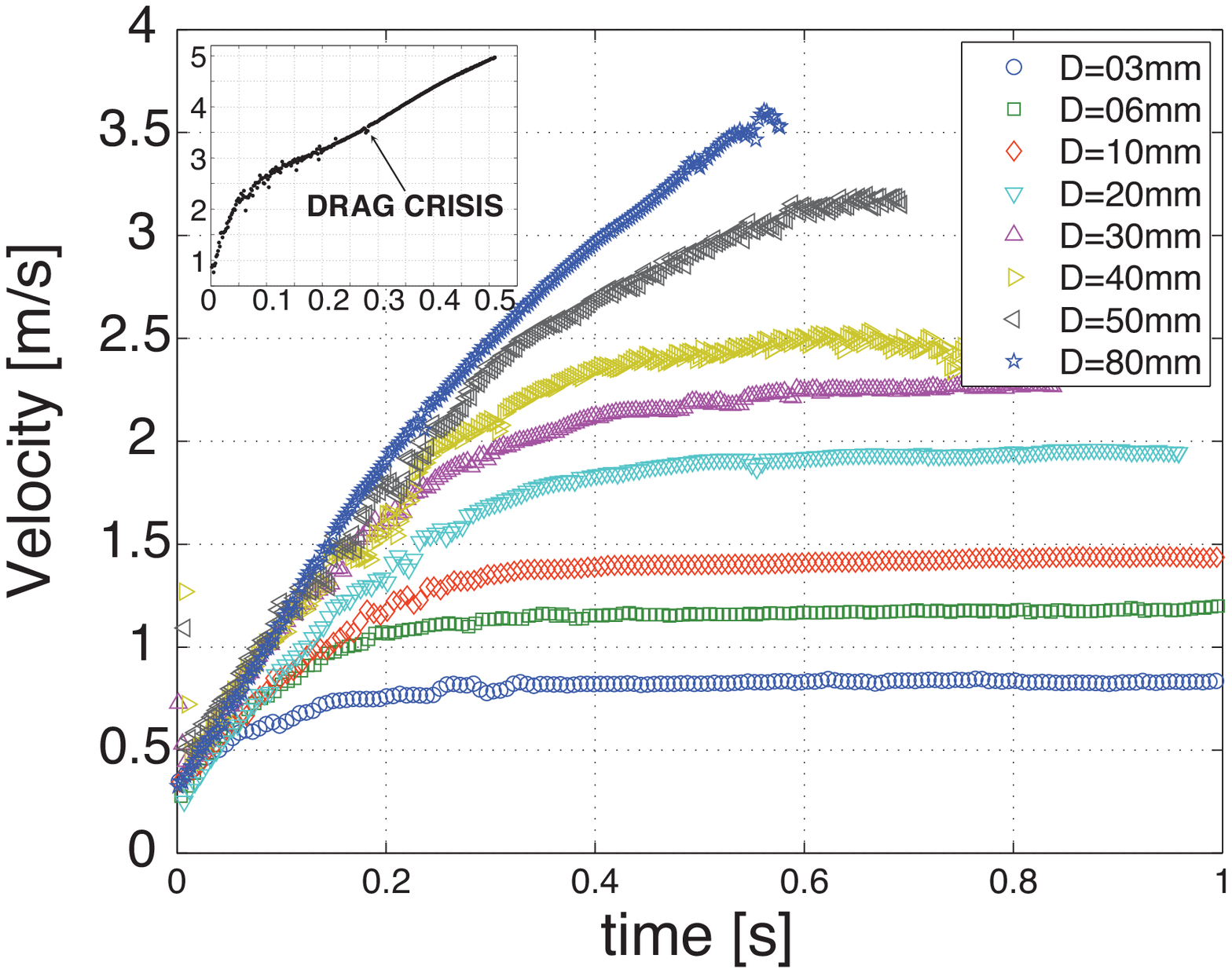}}
\centerline{\includegraphics[width=7.5cm]{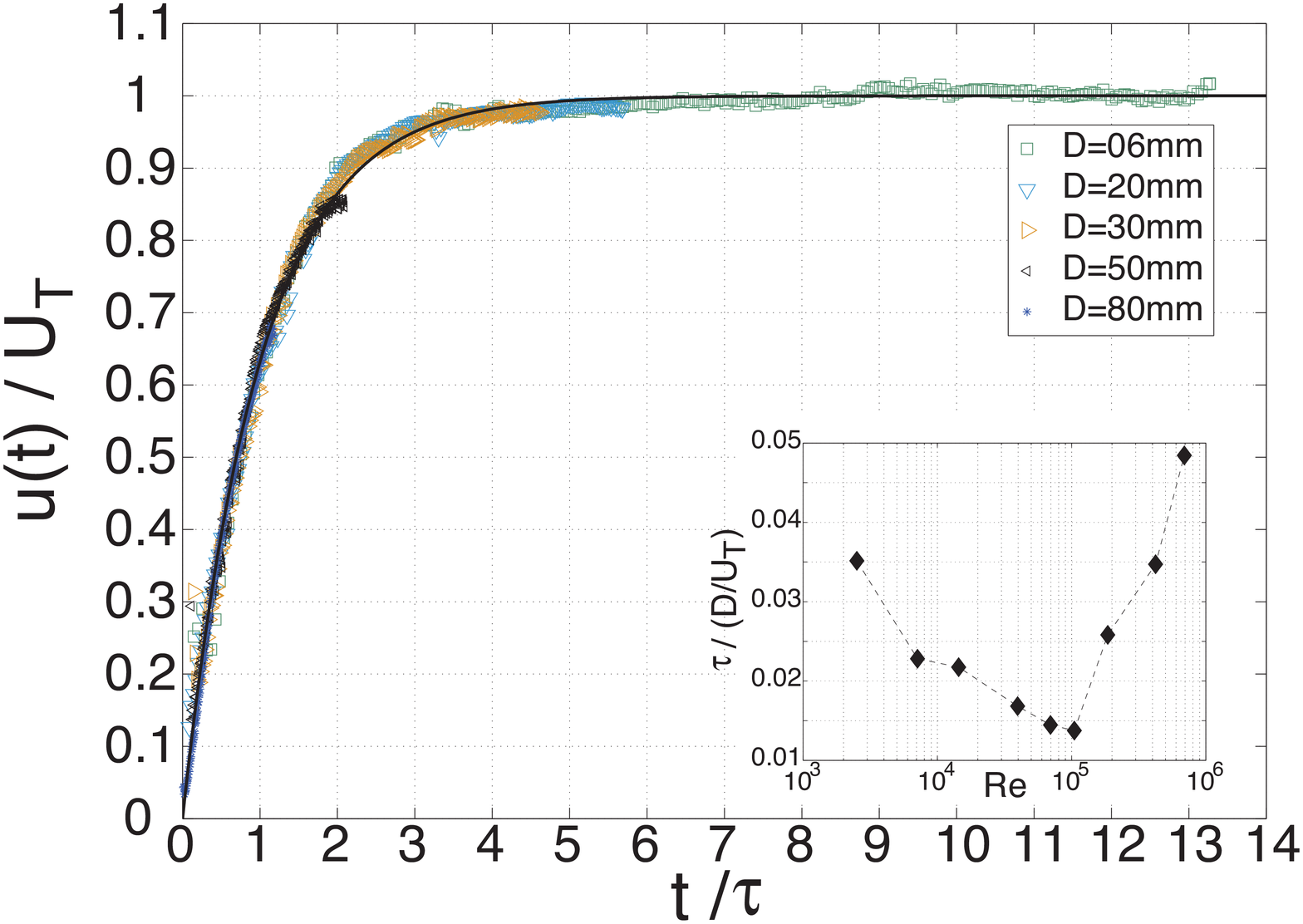}}
 \caption{WATER \& SMOOTH SPHERES: (a) Time series of the spheres vertical velocity $u(t)$ during their free fall. The inset shows the drop -- with a non zero initial velocity -- of a 60~mm sphere: as its velocity reaches 3.5~m/s it meets the drag crisis. (b) Comparison of the experimental data with a model  $u(t) = U_T(1-\exp(-t/\tau))$ exponential evolution. The parameters $(\tau, U_T)$ are obtained using a multidimensional unconstrained nonlinear minimization (Nelder-Mead) with MATLAB. The inset shows the evolution of the characteristic time $\tau$ with the Reynolds number. Note the sharp change in behavior near the drag crisis.}\label{0ppmts}
\end{figure}

\subsection{Water}
We show in fig.~\ref{0ppmts}(a) the fall velocity time series for
the spheres with diameter $D$ varying between 6~mm and 80~mm. As
the spheres are released from rest, they accelerate until a
terminal velocity $U_T$ is reached --  although for the larger
spheres the water tank is not sufficiently tall for this steady
state to be reached. The dynamics at the onset of motion is quite
complex. Added mass effects, as well as wake-induced lift forces
and history forces play a role~\cite{maxeyriley,mordant1}. However
when the Reynolds number is large the dominant forces at work
during the vertical fall of the sphere are the gravitational force
$F_B=1/6 (\rho_S-\rho_F) \pi D^3 g$ and an effective drag force
$F_D=1/8 C_D \pi \rho_F D^2 U_T^2$, where $C_D$ is the usual drag
coefficient; $\rho_F$ and $\rho_S$ are the fluid and sphere
densities. In the steady state, these forces balance and one may
then compute the drag coefficient as
\begin{equation}
C_D=\frac{4}{3}\frac{(\rho_S/\rho_F-1)D g}{U_T^2} \ .
\end{equation}
We note here that, unlike wind tunnel experiments, the velocity is
not prescribed so that both $C_D$ and $Re$ are empirically
computed from the data -- the equation above may be viewed as an
implicit relationship for $C_D(Re) Re^2$ as a function of the
control parameters of the experiment.

We observe that during the approach to terminal speed, the
trajectories for different sphere sizes are fully characterized by
one time scale $\tau$ and the terminal speed $U_T$.  We may
extract $\tau$ and $U_T$ from each velocity time series by fitting
the data to an exponential of the form, $u(t) = U_T(1 -
\mathrm{e}^{-t/\tau})$.  In agreement with previous
observations~\cite{mordant1}, the exponential is simply an
effective tool used to extract $\tau$ and $U_T$ and does not
accurately represent the more complex dynamics of the true
trajectory.  When scaled by $\tau$ and $U_T$, all the time series
in Fig.~\ref{0ppmts}(a) collapse onto one curve, verifying the
importance of these two characteristic quantities.  Using the
exponential fit on the entire time series, we take advantage of
our good resolution in both time and velocity magnitude to obtain
accurate measurements of $U_T$ even though the fall distance is
only 2~m. Since this method integrates the whole time record of
the fall, it also avoids possible errors incurred by taking single
point measurements as has been done in past studies.  Furthermore,
inspection of the entire velocity time series is often very
instructive, clearly revealing the drag crisis in some cases --
see for instance the inset of fig.~\ref{0ppmts}(a), where a 60~mm
sphere is shown to accelerate again as its Reynolds number exceeds
$Re^*_{\rm w}$. 

\begin{figure}
\center
\centerline{\includegraphics[width=8cm]{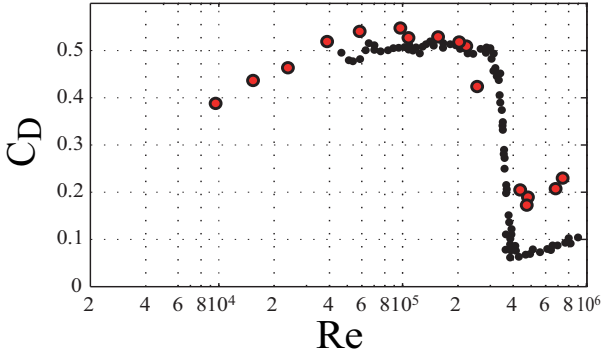}}
\caption{DRAG COEFFICIENT measurements for smooth spheres in water. Red Circles -- our data; solid circles -- Achenbach (wind tunnel)~\cite{achenbach2}. 
}\label{comp0ppm}
\end{figure}

We compare in fig.~\ref{comp0ppm} our measurements
of drag coefficients for smooth spheres in water to the free fall
measurements of A. White \cite{awhite} as well as the wind tunnel
measurements of Achenbach \cite{achenbach1}. We find an excellent
agreement with White's data. In particular, we find that the
critical Reynolds for the drag crisis is $Re^*_{\rm w} = 2.8 \,
10^5$, a value that serves as a reference for comparison with the
fall of spheres with modified surfaces and in water with
additives. We also note that both White's data and ours suggest
that the value of the drag coefficient just after the drag crisis
for the free fall of spheres (imposed force case) is twice that
observed in wind tunnel experiments (imposed velocity case).

\subsection{Polymer solution}\label{polsol}
\begin{figure}
\center
\centerline{\includegraphics[width=8cm]{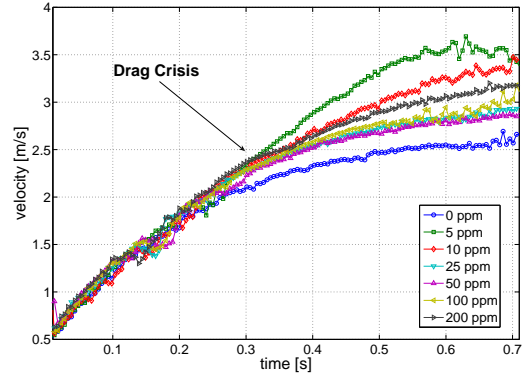}}
\centerline{\includegraphics[width=8cm]{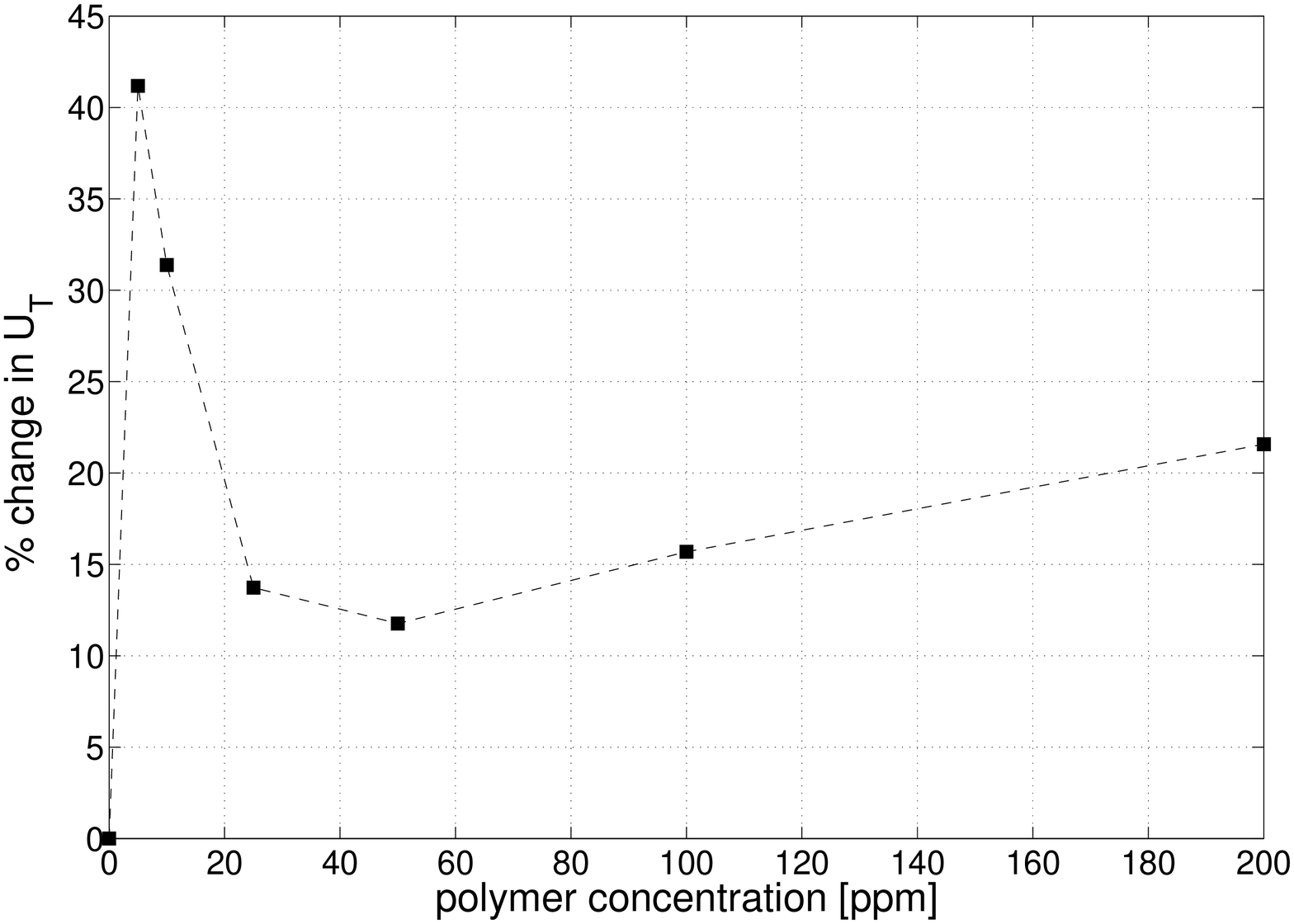}}
\caption{POLYMERS : (a) Fall velocity time series of a 40~mm sphere in water,
and polymer solutions with concentration increasing from 5 to 200
ppm.  In the 5 and 10 ppm solutions the sphere undergoes a drag
crisis where none existed for the pure water case. (b) Percentage
change of terminal velocity $U_T$ for increasing polymer
concentration compared to pure water case for 40~mm sphere.
 }\label{crisis}
\end{figure}

\begin{figure}
\center \centerline{\includegraphics[width=9cm]{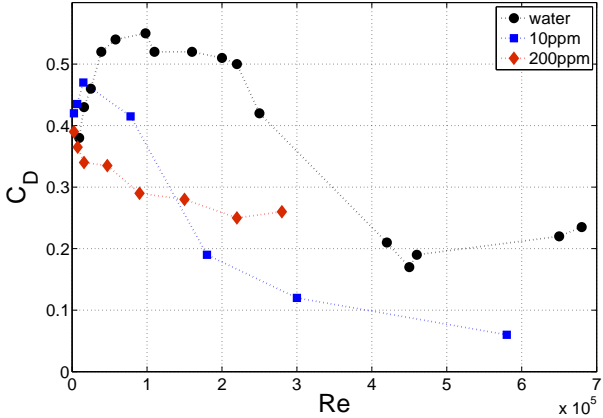}}
\caption{
POLYMERS : Drag coefficient measurements for smooth
spheres in water (solid circles) compared to polymer solution
(open circles). (a) In 10 ppm solution the drag crisis is shifted
to lower $Re$. (b) In 200 ppm solution the drag crisis is replaced
by a gradual decrease in drag.}\label{ppm}
\end{figure}

We first present velocity time series for a 40~mm sphere falling
in a range of polymer concentrations in fig.~\ref{crisis}(a). We
observe that at all concentrations the sphere terminal velocity is
larger than in pure water; drag is reduced.  This effect is
greatest at small polymer concentrations, as demonstrated in
fig.~\ref{crisis}(b): drag reductions over 30\% have been observed
for polymer concentration less than 20 ppm, while at higher
concentrations the change is 10-25\%. In the 5 and 10 ppm
solutions, one observes a sudden acceleration of the sphere once
it achieves a velocity of about 2.5~m/s; this is the drag crisis.
Examining the data for a range of sphere sizes in the 10 ppm
solution (see Fig.~\ref{ppm}(a)), we see that the critical
Reynolds number is then $Re^*_{\rm polymer} \sim 1.0 \, 10^5$,
almost a third of the value $Re^*_{\rm w} \sim 2.8 \, 10^5$ in pure
water. On the other hand, at higher polymer concentrations, we do
not observe a jump in the velocity time series and their is no
discontinuity in the drag $C_D(Re)$ curve.  One observes that for
high polymer concentrations, the drag is reduced at $Re <
Re^*_{\rm w}$, but enhanced for $Re > Re*$: in pure water a drag
crisis would have occured and dramatically reduced $C_D$ but this
does not happen when the polymer concentration exceeds about
100ppm as shown in Fig.~\ref{ppm}(b). Instead the value of drag
decreases continuously. These observations are consistent with the
experiments of previous investigations using poly(ethylene oxide)
in a similar range of concentrations \cite{awhite,watanabe}. We
have not been able to reach Reynolds numbers high enough to
determine whether the drag would reach a common asymptotic limit.

\subsection{Rough and grooved surfaces in water}
\begin{figure}[h!]
\center
\centerline{\includegraphics[width=10cm]{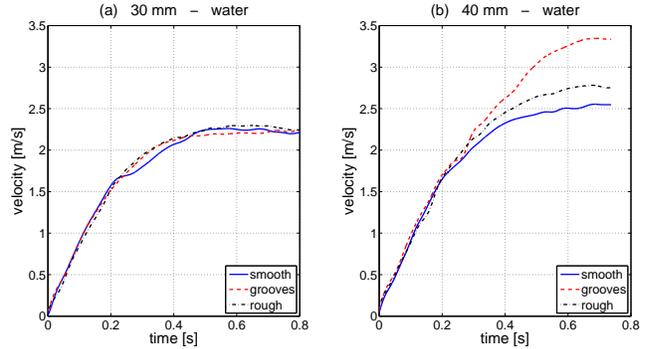}}
\caption{ ROUGH \& GROOVED SPHERES: velocity time series for grooved (dashed line) and rough spheres (dahs-dotted line) compared to smooth spheres (solid line) in pure water.  The grooved surface induces an early drag crisis.}\label{surf0}
\end{figure}

In exploring surface structure effects, we concentrate our
attention on 30 and 40~mm spheres, whose $Re$ in pure water lies
just below the drag crisis.  The time series in fig.~\ref{surf0}
illustrate the different behaviors for the different surfaces we
studied.  In pure water, both the 30~mm grooved sphere and rough
sphere behaves the same as the 30~mm smooth sphere --- cf.
Fig.~\ref{surf0}(a).  In contrast, adding grooves to the 40~mm
sphere induces a drag crisis, as shown in Fig.~\ref{surf0}(b). The
40~mm rough sphere showed moderate drag reduction, but not a well
defined crisis. Indeed, the dynamics in Fig.~\ref{surf0}(c) shows
that the terminal velocity is increased compared to the smooth
sphere, but there is no clear change in the acceleration as in the
case of the grooved sphere, Fig.~\ref{surf0}(b).

Grooves are thus able to shift the drag crisis from $Re^*_{\rm w} \sim 2.8 \, 10^5$ to $Re^*_{\rm grooves}  \sim 0.8 \, 10^{5}$. In the case of the 40~mm sphere, the terminal velocity increases from 2.5~m/s to 3.4~m/s, corresponding to a drag reduction of 46\%. For the rough sphere, a drag reduction is also observed but it is limited to a 20\% gain. This difference in behavior is not yet understood. One may note that a rough surface destabilizes the boundary layer but also increases friction and dissipation. 

Finally, we have observed that sanded spheres (rugosity of the order of 10 $\mu$m) with a diamater of 30 and 40~mm showed no change compared to smooth spheres. This indicates that surfaces modifications must exceed the thickness of the viscous sub-layer in order to produced measurable effects on the dynamics.

\subsection{Rough and grooved surfaces in polymer solution}
\begin{figure}
\centerline{\includegraphics[width=10cm]{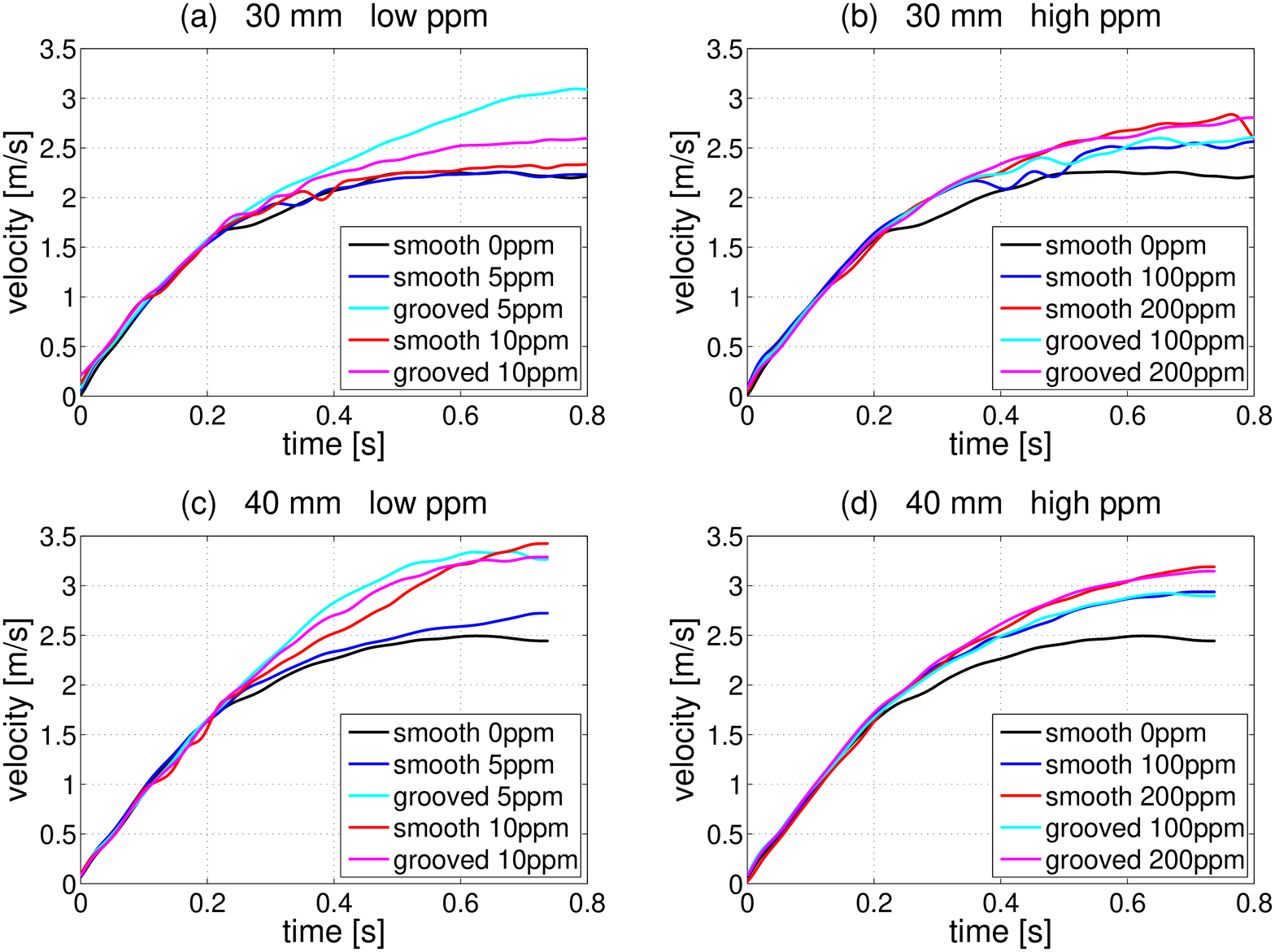}}
\caption{GROOVED SPHERES \& POLYMERS: At low polymer concentration (left
column) adding grooves to the sphere induces an even earlier drag crisis compared to the smooth sphere.  At high concentration (right column) grooves do not change the observed dynamics.}
\label{surfpoly2}
\end{figure}

We now examine the changes in the above described behavior when polymer is added to the water.  We find that the two regimes of low and high concentration --  section~\ref{polsol}  -- are affected differently by adding grooves to the sphere surface. Results for the  grooved spheres are presented in fig.~\ref{surfpoly2}.  At low concentration the shift of the drag crisis to lower $Re$ due to polymer is exaggerated by adding grooves to the sphere; $Re^*_{\rm w}$ is shifted even lower. Indeed, in a 5 ppm solution, the 30~mm grooved sphere experiences the drag crisis, whereas the same sphere in water as well as the smooth 30~mm sphere in 5 ppm solution do not. We find that the $Re^*_{\rm grooved + poly} \sim 6 \, 10^{4}$, a further gain of 20\% compared to polymers alone. The same behavior is observed for the 40~mm grooved sphere at low polymer concentration.   
At higher polymer concentration, the spheres behave identically with or without grooves.

The rough sphere did not exhibit the same behavior. Rather, the surface roughness seems to suppress the drag crisis, in agreement with the observations of A. White \cite{awhite2}.  Our results are presented in Fig.8, for a 40~mm sphere. When the surface is smooth, one observes as before the shift in the drag crisis and a
very large terminal velocity at low polymer concentration (10~ppm), as well as a reduced drag at high concentration (200~ppm). However, for the rough sphere all dynamical $v(t)$ curves are very close. The rough spheres experience no further decrease in drag in the polymer solutions, compared to what is already induced by the surface roughness.  In fact, there even may be a slight increase in drag (of the order of 5\%)  when the rough sphere falls in the water and polymer solution, at any concentration.

\begin{figure}
\centerline{\includegraphics[width=8cm]{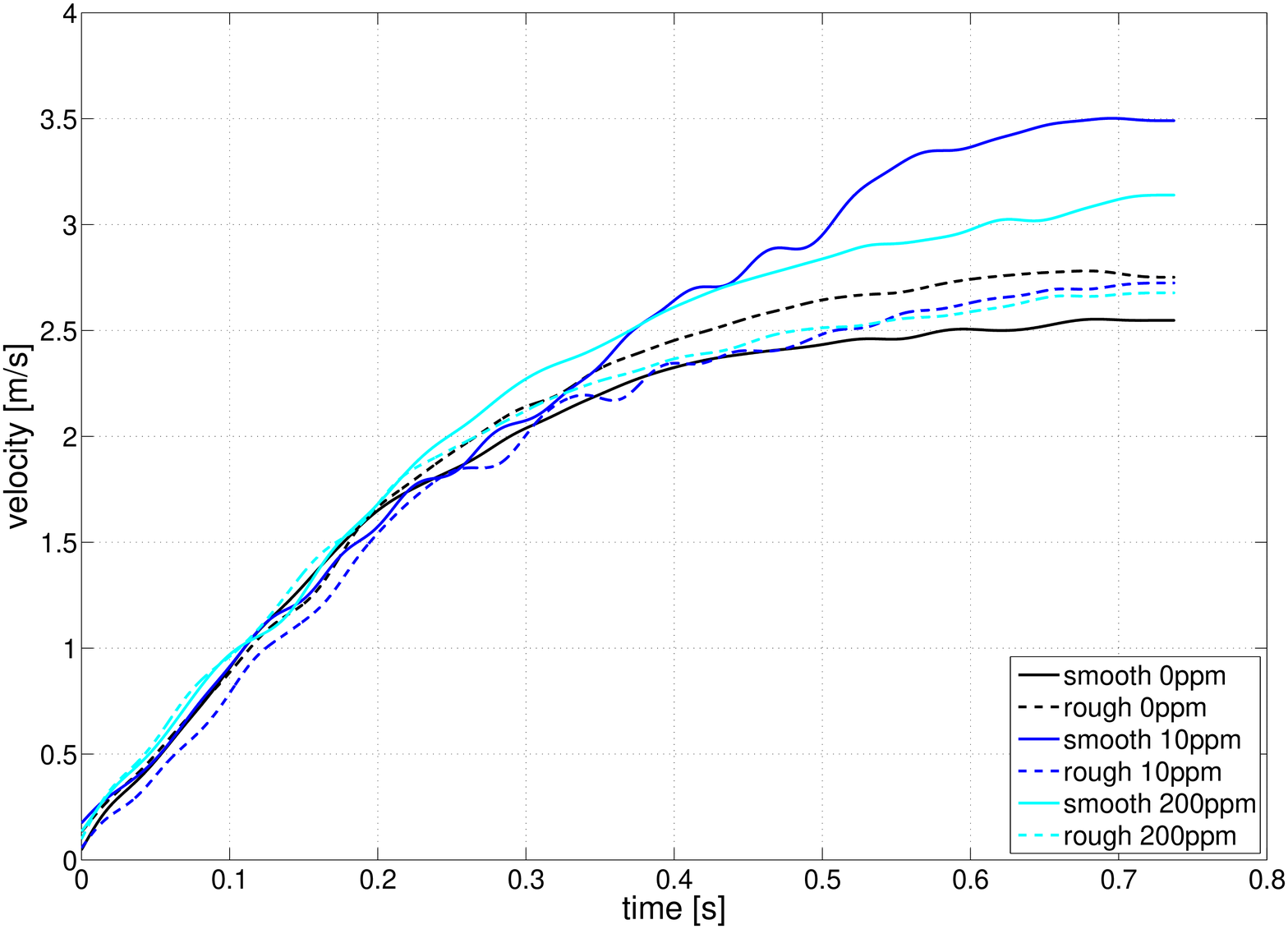}}
\caption{ROUGH SPHERE: Adding polymer causes nearly no change in
the behavior of rough spheres apparently suppressing the drag
crisis independent of the polymer concentration.
}
\label{roughpoly}
\end{figure}

\section{Discussion}
\subsection{Experimental summary}
We have conducted a series of experiments using precise and time resolved ultrasound velocity measurements to compare the behavior of rough and smooth steel spheres falling through water or dilute aqueous polymer solutions. Remarkably, we find that in low concentration polymer solutions (5 to 20 ppm) the drag crisis
happens at a lower Reynolds number than in water.  By adding a pattern of shallow grooves to the sphere surface, we shift the drag crisis to even lower $Re$.  Adding grooves to a sphere in pure water also shifts the drag crisis to lower $Re$.  On the
other hand, a sphere roughened with a layer of 700~$\mu$m beads glued to its surface never experiences a drag crisis, exhibiting nearly the same drag with or without polymers. The drag on a rough sphere is slightly less than that on a smooth sphere.  For higher concentration polymer solutions (100 - 200 ppm) and smooth
spheres the drag crisis is suppressed and replaced by a more gradual decrease in drag as $Re$ is increased. This high concentration behavior is largely unchanged by adding grooves to the sphere surface. \\

Our measurements seem to indicate that for low concentrations the polymers are able to induce the transition to turbulence but have little effect on the location of flow separation whether laminar or turbulent.  That is, low polymer concentrations induce an early drag crisis, but do not greatly change the drag before and after the crisis, so that we may conclude that the location of the separation points have not been significantly changed. In fact, we have observed that the dynamical behavior $v(t)$ is quite well modelled by a simple shift in the $C_D(Re)$ curve, coupled to a simple dynamical equation in which only the drag force is accounted for.

At high concentration and $Re<Re^*_{\rm w}$ (i.e. laminar flow separation), drag is reduced, which implies that the separation location is pushed downstream on the sphere surface. On the other hand, for the case of turbulent flow separation ($Re>Re^*_{\rm w}$), $\theta_s$ apparently shifts upstream, which manifests as an increase in drag.

Surface roughness is commonly understood to induce an early transition to boundary layer turbulence \cite{maxworthy}, which may explain the shift in $Re^*_{\rm w}$ observed for the grooved sphere. On the other hand, it is difficult to explain in the same context our observation of rather weak drag reduction and apparent
suppression of the drag crisis for the rough sphere. Perhaps friction drag is significant in this case. Further investigation of this curious behavior is left for future work.

\subsection{Drag crisis and normal stress difference}\label{lyderic}
In this section we try to rationalize the effect of the polymers on the observed drag reduction. We follow ideas proposed for drag reduction in pipes~\cite{Lumley2} and much devloped since (see for instance~\cite{dragred}). Specifically, a change of conformation of the polymer is argued to be the source of the modification of the drag crisis. 

As discussed above, the drag crisis is the result of the destabilization of the laminar boundary layer \cite{Schlichting}. At a critical Reynolds number 
the boundary layer becomes turbulent, shifting the separation line downstream and reducing accordingly the drag on the sphere. The polymer has {\it a priori}Ê little effect on the parameters influencing this boundary layer transition, like the viscosity $\eta$. Indeed the polymer concentration is smaller than the overlap concentration $\xi^\star$, separating the dilute from the semi-dilute regime \cite{Doi} -- for the polymers under consideration, this is estimated to be $\xi^\star\simeq1200$~ppm. The shear viscosity of the polymer solutions in water, $\eta_P$,  is related to the polymer density according to $\eta_P=\eta_w(1+1.49 {\xi / \xi^\star})$ with $\eta_w$ the water viscosity \cite{Doi}. Thus for the low concentrations under consideration here,
$\xi\ll\xi^\star$, the viscosity is close to that of water $\eta \sim \eta_w$.

\begin{figure}[t]
\centerline{\includegraphics[width=8cm]{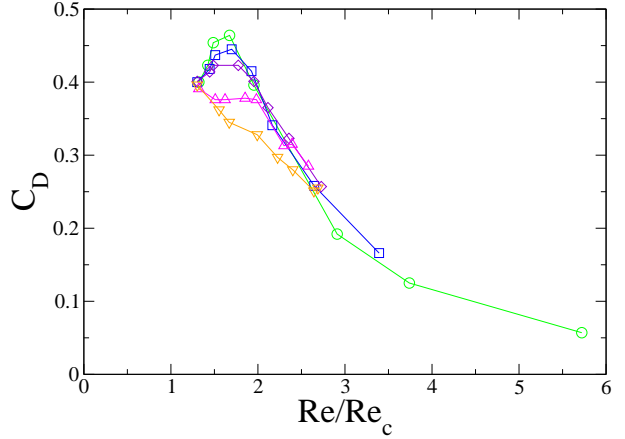}}
\caption{Drag coefficient versus the reduced Reynolds number
$Re/Re_c$. The different symbols correspond to various
concentrations of polymers: ($\circ$) 10ppm; (square) 25ppm;
($\diamond$) 50ppm; ($\bigtriangleup$) 100ppm;
($\bigtriangledown$) 200ppm. Lines are a guide to the eye. For a
given polymer concentration, the different experimental points
correspond to different size of the falling sphere (from left to
right, $D=3,6,10,20,30,40,50,60$mm). }
\label{figlyd}
\end{figure}

However this estimate assumes that the polymers' structure is not affected by the flow. Velocity gradients may locally induce a stretching of the polymer, which can be quantified by the Weissenberg number defined as $Wi=\dot\gamma\tau_R$, with $\dot\gamma$ a deformation rate and  $\tau_R$ the polymer relaxation time. Typically, for $Wi<1$ the polymer is in a coil state, while for $Wi >1$ stetching occurs. Let us estimate  $Wi$ in our geometry. The relaxation time is typically $\tau_R\sim {\eta_w R_g^3 / k_BT}$, with $R_g$ the radius of gyration of the polymer, $R_g\sim b N^{\nu_F}$ ($b$ the monomer size and $\nu_F\simeq 3/5$ the Flory exponent). For the polymers under investigation, $\tau_R\sim 10^{-4} s$. On the other
hand the deformation rate is estimated as the shear rate in the boundary layer, {\it i.e.}Ê $\dot\gamma\sim U/\delta$, with $U$ the sphere velocity and $\delta\sim\sqrt{\nu D/U}$ the typical thickness of the boundary layer ($a$ the diameter of the sphere).
This gives 
\begin{equation}
Wi \sim {U^{3/2}\tau_R\over \sqrt{\nu D}} \ , 
\end{equation}
which can be rewritten
\begin{equation}
Wi\sim \left({Re\over Re_c }\right)^{3/2} \ ,
\end{equation}
with a critical Reynolds number $Re_c$ defined as
\begin{equation}
Re_c= \left( {D^2\over\nu\tau_R }\right)^{2/3} \ .
\end{equation}
At $Re_c$ the polymer is thus expected to undergo - within the boundary layer - a coil-stretched transition  and the drag will be accordingly be affected (as we discuss hereafter). This point is confirmed experimentally in Fig. \ref{figlyd} where the drag
coefficient is plotted versus the reduced Reynolds number $Re/Re_c$ : the 'drag-crisis' is always found to occur for $Re\sim Re_c$ for the different cases investigated. While a full rescaling is not expected in this plot, this figure points to the relevance
of the Weissenberg number as a key parameter to the polymer induced drag crisis: it does show that the drag crisis transition with polymers, {\it i.e.} when the drag coefficient strongly decreases, occurs at a Reynolds number of the order of the
critical Reynolds number, $Re\sim Re_c$. This indicates that the drag crisis criterion with polymers corresponds to $WiÊ\sim 1$, as also observed in earlier works.\\

At this level, the previous discussion suggests that the polymer effect on the drag crisis is associated with a conformation change. The question of the polymer-flow coupling however remains, and in particular the origin of an earlier destabilisation of the boundary layer.

First, as the polymers in the boundary layer become stretched, most of the properties of the polymer solution in this region will change dramatically : the typical size of the polymer increases indeed from the radius of gyration to the much larger contour
length of the polymer, $L \gg R_g$. This affects the relaxation time which now becomes $\tau_R \sim {\eta_w L^3 / k_BT}$, and therefore the viscosity which increases typically by a factor $(L/R_g)^3= N^{3(1-\nu_F)}\gg 1$. However, increasing the
viscosity in the boundary layer amounts to a {\it decrease} in the local Reynolds number : this would lead to a re-stabilisation of the laminar boundary layer, an effect which is opposite to the experimental observation.

Another origin has therefore to be found. We suggest that the destabilisation of the boundary layer originates in a very large {\it normal stress difference} occuring when the polymer is in its strechted state. Normal stress difference is a non-newtonian
effect which is commonly observed in polymeric solutions \cite{Doi}. This is known to lead for example to the Weissenberg (rod-climbing) effect. In our geometry, the normal stress difference is expected to be proportional to the square of the
shear-rate according to
\begin{equation}
\Delta\sigma=\sigma_{xx}-\sigma_{yy}=\Psi_P \dot\gamma^2
\end{equation}
with $\Psi_P$ a transport coefficient ; $\sigma_{xx},\sigma_{yy}$ are the normal components of the stress tensor in the $x$ and $y$ directions, with $\{x,y\}$ local coordinates respectively parallel and perpendicular to the sphere surface (curvature effects are neglected).

Let us show that this term does destabilize the boundary layer. Classically, the boundary layer is destabilized by a {\it negative} pressure gradient term due to a decrease of the fluid velocity $U_e(x)$ in the outer layer \cite{Schlichting} : $-\nabla
P_e=\rho U_e(x)\nabla U_e(x)$, with $U_e(x)$ the fluid velocity outside the boundary layer. A stability analysis of the boundary layer with such a pressure gradient leads to a destabilization at a reduced Reynolds number\cite{Schlichting} $Re_\delta=U\delta/\nu \sim 600$, corresponding to $Re\sim 10^5$. The normal stress difference adds a contribution to this term, leading to an supplementary effective pressure gradient term 
\begin{equation}
-\nabla P_{\rm eff}=\rho U_e(x)\nabla U_e(x)+\Psi_P
\nabla\dot\gamma^2,
\end{equation}
where $\dot\gamma\simeq U_e(x)/\delta(x)$ and $\delta(x)\simeq \sqrt{\nu x/U_e(x)}$ the local thickness of the boundary layer. It is easy to verify that this supplementary contribution to the effective pressure gradient will be negative - and therefore destabilizing -, {\it before} the classical contribution $\rho U_e(x)\nabla U_e(x)$. Moreover in the stretched state - for $Wi>1$-, one may verify that this contribution is dominant as compared to the classical one. The ratio $\Delta$ between these
two terms is of order $\Delta\sim\Psi_P \dot\gamma^2/\rho U_e^2$.
Using $\Psi_P\sim \eta_P \tau_P$ with $\eta_P$ the polymer
contribution to the viscosity and $\tau_P$ the polymer relaxation
time \cite{Doi}, one deduces $\Delta\sim U\tau_P/D \sim (L/R_g)^3
/\sqrt{Re_c}$ (for $Re=Re_c$). In our case, with $Re_c\sim 10^5$,
$(L/R_g)^3=N^{3(1-\nu_F)}\sim 2. 10^5$ ($N\simeq 35.10^3$), one
has $\Delta\sim 10^3\gg1$. This term thus leads to a strong destabilization as soon as the polymer is stretched.\\

To summarize, for $Re\ge Re_c$ the coil-stretched transition
occurs for the polymer in the boundary layer, and the existence of
a normal stress difference induces a strong destabilization of the
laminar boundary layer. This scenario gives the trends of the underlying mechanisms
leading to a shift of the drag crisis even for very small amounts
of polymers. For the polymer additive to have an effect, the
critical Reynolds number has to be {\it lower} than the critical
Reynolds number for the drag crisis in pure water, $Re_w^\star$:
$Re_c= \left( {D^2\over\nu\tau_R }\right)^{2/3}< Re_w^\star$. This
provides a condition in terms of the size of the falling object
but also a minimal polymer weight (since $\tau_R\propto
N^{3\nu}$). To go further, a more detailed stability analysis of the boundary
layer with the supplementary normal stress difference is needed. We leave this point for further studies.


\end{document}